\documentclass[aps,prb,twocolumn,superscriptaddress,10pt]{revtex4}
\usepackage[dvips]{graphics}
\usepackage{graphicx}
\usepackage{amsfonts}
\usepackage{amssymb}
\usepackage{amsmath}
\usepackage{color}
\usepackage{hyperref}

\definecolor{applegreen}{rgb}{0.55, 0.71, 0.0}

\begin{document}

\title{From Nodeless States and Vortices to Gray Rings and Symmetry-Broken States in Two-Dimensional Polariton Condensates}

\author{A.S.\ Rodrigues}
\affiliation{Departamento de F\'{\i}sica/CFP, Faculdade de Ci\^{e}ncias, Universidade do Porto, R. Campo Alegre,
687 - 4169-007 Porto, Portugal}

\author{P.G.\ Kevrekidis}
\affiliation{Department of Mathematics and Statistics, University of Massachusetts,
Amherst MA 01003-4515, USA}

\author{R.\ Carretero-Gonz\'{a}lez}
\affiliation{Nonlinear Dynamical Systems Group,%
\footnote{\texttt{URL}: http://nlds.sdsu.edu}%
Department of Mathematics and Statistics, and Computational Science
Research Center, San Diego State University, San Diego CA, 92182-7720, USA}

\author{J.\ Cuevas}
\affiliation{Grupo de F\'{\i}sica No Lineal.  Departamento de F\'{\i}sica Aplicada I.
Escuela Polit\'ecnica Superior, Universidad de Sevilla, C/ Virgen de \'Africa, 7, 41011-Sevilla, Spain}

\author{D.J.\ Frantzeskakis}
\affiliation{Department of Physics, University of Athens, Panepistimiopolis,
Zografos, Athens 157 84, Greece}

\author{F.\ Palmero}
\affiliation{Grupo de F\'{\i}sica No Lineal.  Departamento de F\'{\i}sica Aplicada I.
Escuela T\'ecnica Superior de Ingenier\'{\i}a Inform\'atica, Universidad de Sevilla, Avda. Reina Mercedes, s/n, 41012-Sevilla, Spain}

\begin{abstract}
We consider the existence, stability and dynamics
of the nodeless state and fundamental
nonlinear excitations, such as vortices,
for a quasi-two-dimensional polariton condensate
in the presence of pumping and nonlinear damping.
We find a series of interesting features that can be directly contrasted
to the case of the typically
energy-conserving ultracold alkali-atom
Bose-Einstein condensates (BECs). For sizeable parameter ranges,
in line with earlier findings,
the nodeless state becomes unstable
towards the formation of {\em stable} nonlinear single or multi-vortex
excitations. The potential instability of the single vortex is also
examined and is found to possess similar characteristics to those
of the nodeless cloud.
We also report that, contrary to what is known, e.g., for the atomic
BEC case, {\it stable} stationary
gray rings (that can be thought of as radial forms of a Nozaki-Bekki
hole) can be found  for polariton condensates
in suitable parametric regimes. In other regimes, however, these
may also suffer symmetry breaking instabilities.
The dynamical, pattern-forming implications of the above instabilities
are explored through direct numerical simulations and, in turn, give rise
to waveforms with triangular or quadrupolar symmetry.
\end{abstract}
\date{\today}

\maketitle

\section{Introduction}

One of the most rapidly developing branch of studies in the physics
of Bose-Einstein condensation
is that of exciton-polariton condensates in semiconductor microcavities.
Only a few years since their experimental
realization~\cite{kasp1_and_more,balili,lai,deng},
exciton-polariton Bose-Einstein condensates (BECs) have become a prototypical system for studies at
the interface of non-equilibrium physics and nonlinear dynamics.
More specifically, the radiative lifetime
of the polaritons provides a short relaxation time scale in the system
of the order of 1--10~ps~\cite{benoit}. At the same time, the light mass
of these quasi-particles provides them with a considerably higher
condensation temperature. Moreover, the photonic component of the system
only allows for a short lifetime and no thermalization. Instead, the
exciton-polariton system produces a genuinely non-equilibrium condensate,
requiring the external pumping of an excitonic reservoir, which in turn
balances the polariton loss~\cite{benoit,sanvitto}. This ``open'' nature
of the system, featuring gain and loss, is then responsible for its
rich pattern forming capabilities that have been recently summarized,
e.g., in Refs~\citenum{rmp1,berloff_review,rmp2}.

Our interest in the prototypical two-dimensional setting of the
system will be precisely in the interplay of the intrinsic nonlinearity
due to inter-particle interactions and the gain/loss nature of the
system. This interplay has led to a wide variety of remarkable
observations (and theoretical explorations) including, but certainly
not limited to, features such as flow without scattering (analogue of the
flow without friction) \cite{amo1}, the existence of vortices \cite{lagou1}
(see also Ref.~\citenum{roumpos} for vortex dipole dynamics
and Ref.~\citenum{roumpos1} for observations thereof),
persistent currents as well as higher charge vortices~\cite{sanvitto_nat},
collective dynamics \cite{amo2},
solitary wave structures such
as bright~\cite{skryabin},
dark~\cite{dev} and gap~\cite{amogap} solitons,
and even remarkable applications
such as spin switches \cite{amo3} and light emitting diodes \cite{amo4}
operating even near room temperatures.

The approach that has been used most commonly in
theoretical studies
of exciton-polariton BECs relies on the analysis of two coupled evolution
equations for
the polaritons and
the exciton reservoir which enables their production. In particular, the relevant
model assumes the form of two
coupled complex Ginzburg-Landau (cGL) equations describing the evolution
of exciton and photon wavefunctions~\cite{polar1,polar2,cc05}.
However, an alternative that has been proposed~\cite{berloff1,kbb,kbb2,rmp2}
in the case of incoherent and/or far blue-detuned laser pumping (see, e.g.,
Ref.~\citenum{rmp2} and references therein)
suggests that a single cGL equation
for the macroscopically occupied polariton state
may be used instead; such a model yields results consistent with
experimental observations~\cite{nbb_nature} (see also Refs.~\citenum{berloff_review,rmp2}).

In what follows, we will consider the case of incoherent pumping and use in our study
a single cGL equation.
Our aim 
is to analyze in detail
some of the
fundamental states of the two-dimensional system. In particular,
in earlier works these states have been chiefly obtained as
attractors of the relevant gain/loss dynamics, revealing the pattern
forming complexity that emerges spontaneously in the system.
Here, our aim is not only to revisit fundamental states
(such as the nodeless cloud or the single vortex) and explore their
parametric dependence by developing two-parameter bifurcation diagrams
(in parameters such as the gain strength and its spot radius);
it is instead to provide a detailed view towards the stability of
these states unveiling their spectral properties and
the somewhat unusual nature of their instabilities. In addition
to these more standard states, we will also
consider states that, to the best
of our knowledge, have not been previously presented
in the context of polariton condensates,
although they have been discussed for atomic condensates~\cite{dark_djf}.
A principal example of this form is the so-called ring dark soliton (RDS) which, remarkably,
although never stable in the context of atomic
BECs~\cite{herring,todd,theoch_1}, can in fact be shown to be stable in
suitable (gain) parametric regimes here. This is, effectively, a potentially
stable radial form of a Nozaki-Bekki hole~\cite{nozakibekki} that was previously explored in
cGL contexts~\cite{rabin1}, yet was not found to be stable in
these settings; instead, it was found there to 
potentially initiate a form of spiral wave turbulence.
We also reveal the
symmetry-breaking instabilities
of this ring structure and
unveil a series of solutions without radial symmetry that may spontaneously
emerge as a result of such instabilities. Among them, we highlight the potential
for states with triangular or square/rectangular symmetry, whose parametric dependence (as stationary
states) we also explore.
Finally, for all relevant states, we offer a number of direct numerical
simulations that yield insight towards the manifestation of the instabilities
and the spontaneous emergence of patterns such as vortex lattices, but also
of non-vortical patterns without radial symmetry. We should also note
in passing here that a similar study focusing, however, predominantly
on the existence properties of some of the solutions considered here
(rather than on their stability, which is the
principal emphasis herein), and chiefly considering the case without
a parabolic trap, was recently published~\cite{OstAbd12}.

Our exposition is structured as follows. In section II, we offer
the theoretical setup and techniques that will be used. In section III,
we present the numerical results in two subsections: the first one
provides the bifurcation structure and parametric continuations/stability
analysis of the relevant solutions (initially this is done for the nodeless
cloud and single vortex, and subsequently for the ring and related symmetry broken
states); the second one examines the results of
direct numerical results. Finally, in section IV, we summarize our findings,
as well as mention some interesting directions for potential future
studies.

\section{Model setup}

As indicated above, we will consider the complex Ginzburg-Landau
model developed in
Refs.~\citenum{berloff1,kbb,kbb2} (see also Ref.~\citenum{rmp2} and references therein):
\begin{equation}
\label{eq:dyn}
i\partial_t\psi=
\left\{-\nabla_\perp^2+r^2+|\psi|^2+i\left[(\chi(r)-\sigma|\psi|^2)\right]\right\}\psi,
\end{equation}
where $\psi$ denotes the polariton wavefunction trapped inside
a two-dimensional (2D)
harmonic potential, $\nabla_\perp^2 \equiv \partial_x^2+\partial_y^2$ is the transverse (2D) Laplacian
and $r^2 \equiv x^2+y^2$ (note that the $z$-direction
corresponds to the tight trapping axis).
In fact, the above equation has the form of a ``modified'' Gross-Pitaevskii equation (GPE), which
is the traditional lowest-order mean-field model describing atomic BECs \cite{stringari1,stringari2}:
the differences of Eq.~(\ref{eq:dyn}) from the traditional form of the GPE
can be traced in the presence of (i) the spatially dependent gain term with
\begin{equation}
\chi(r)=\alpha\Theta(r_m-|r|),
\end{equation}
where $\Theta$ is the step function generating a symmetric spot
of radius $r_m$ and strength $\alpha$ for the gain,
and (ii) the nonlinear saturation loss term,
of strength $\sigma$.
Estimates of the relevant physical time and space scales, as well as physically relevant
parameter values, are given, e.g., in Ref.~\citenum{berloff1}. It is relevant to
mention that although our results below are given in the
context of Eq.~(\ref{eq:dyn}), we have ensured that similar phenomenology
arises in the model of Refs.~\citenum{polar1,polar2,cc05}, for suitable
parametric choices.
In that light, the phenomenology that is reported in this work
should be {\it broadly} relevant to (2D) polariton BECs independently of model
specifics. We also note in passing that Ginzburg-Landau-type models, similar to
the one of Eq.~(\ref{eq:dyn}) ---i.e., including
a localized gain term (but, in most cases, in the
one-dimensional setting and without the external potential)---
were recently studied in the context of nonlinear optics \cite{boris1} and in the physics of magnon condensates \cite{boris2}.

In what follows, we will consider the
stationary solutions of
this 2D model, in the form
$\psi(r,t) = \psi_0(r)\exp(-i \mu t)$
where $\mu$ is the dimensionless chemical potential, and
the stationary state $\psi_0(r)$ is governed by
the elliptic partial differential equation of the form:
\begin{equation}
\label{eq:stat}
\mu\psi_0=\left\{-
\nabla_\perp^2
+ r^2 +|\psi_0|^2 + i\left[(\chi(r)-\sigma|\psi_0|^2)\right]\right\}\psi_0.
\end{equation}
Importantly, an additional population balance constraint, i.e., an
overall balancing of gain and loss within the 2D domain, has to be enforced:
this condition is $dN/dt=0$, where the norm $N=\int \mathrm{d}^2r |\psi_0|^2$
depicts the number of polaritons. It is straightforward to show that the balance
condition can be readily expressed as:
\begin{equation}
    \int \mathrm{d}^2r\,(\chi(r)-\sigma|\psi_0|^2)|\psi_0|^2=0.
\label{eq:stat1}
\end{equation}
It then follows that the above equation self-consistently selects the particular value of the chemical
potential once the other parameters (i.e., $\alpha$, $\sigma$, and $r_m$)
are fixed.
We note in passing the significant difference of this trait from
the Hamiltonian atomic BEC case, where there exist monoparametric
families of solutions as a function of $\mu$ (which is a free
parameter there rather than one dependent on the remaining gain/loss
parameters).

Once stationary solutions of the differential-algebraic
system of Eqs.~(\ref{eq:stat})-(\ref{eq:stat1}) are identified,
their linear (spectral)
stability is considered by means of a Bogolyubov-de Gennes (BdG)
analysis \cite{stringari1,stringari2}. Specifically,
small perturbations (of order $\mathcal{O}(\delta)$, with $0< \delta \ll 1$)
are introduced in the form
\begin{eqnarray}
\psi(x,y,t)&=&
e^{-i \mu t} \left[\psi_0(x,y) + \delta\, p(x,y,t) \right],
\label{linearization0}
\end{eqnarray}
with
\begin{eqnarray}
p(x,y,t) & \equiv &
a(x,y) e^{-i \omega t} + b^{\star}(x,y) e^{i \omega^{\star} t}.
\label{linearization}
\end{eqnarray}
Then, the ensuing linearized equations
are solved to $\mathcal{O}(\delta)$,
leading to the following eigenvalue problem:
\begin{equation}
    \omega
    \left(\begin{array}{c} a(x,y) \\ b(x,y) \end{array}\right)=
    \left(\begin{array}{cc} L_1 & L_2 \\ -L_2^* & -L_1^* \end{array}\right)
    \left(\begin{array}{c} a(x,y) \\ b(x,y) \end{array}\right),
\end{equation}
for the eigenfrequency $\omega$
and associated eigenvector $(a(r),b(r))^T$, and
$L_1$ and $L_2$ are the following operators:
\begin{eqnarray}
    L_1&\!=\!&-\mu-\frac{d^2}{dx^2}-\frac{d^2}{dy^2}
    +r^2+2(1-i\sigma)|\psi_0|^2+i\chi(r),
\notag
\\[1.0ex]
\notag
    L_2&\!=\!&(1-i\sigma)\psi_0^2.
\end{eqnarray}
When the eigenfrequencies are found to possess a positive imaginary
part, then, per the ansatz of Eq.~(\ref{linearization}), an instability
is expected to arise. On the other hand, if all the spectrum
has Im$(\omega)<0$, then the corresponding structure is spectrally
stable. When a structure is found to be unstable, we conduct
direct numerical simulations of Eq.~(\ref{eq:dyn}) in order
to explore the evolution of the instability and
the state towards which the dynamics is attracted.

We now proceed to study the existence, stability and nonlinear dynamics of
the different configurations of
interest, namely the nodeless cloud, the single-charge vortex and the
ring dark soliton-like waveform, as well as of 
some symmetry-breaking structures
that result from the evolution dynamics of these states, when unstable.

\section{Numerical results}

\subsection{Existence and Spectral Stability}

\subsubsection{Nodeless Cloud and Central Vortex}

We performed a search of nonlinear excitations for different values of the parameters. In what follows,
we chose to keep $\sigma=0.35$ (following the work of
Ref.~\citenum{Cue11}) fixed, and vary both the gain strength, $\alpha$, and the gain spot size radius, $r_m$,
in order to develop two-parameter bifurcation
diagrams characterizing the stability properties of the different states of
interest. The relevant solutions were numerically obtained
by using a (modified) Newton-Raphson method \footnote{The modification of the Newton-Raphson method consists in performing the (singular) Jacobian inversion by means of a LSQR method.} in order to identify (and perform
continuations on) solutions of Eq.~(\ref{eq:stat}),
together with condition~(\ref{eq:stat1}). This system forms a partial
differential algebraic set of equations (PDAE).

We start by exploring the more fundamental solution profiles,
namely the nodeless cloud (NC) and the central vortex cloud (CV).

It would be relevant to recall here, for comparison purposes,
the stability properties of these waveforms in the Hamiltonian
case of $\alpha=\sigma=0$. There, the NC is the
ground state of the system and is neutrally stable for all parameter
values~\cite{stringari1,stringari2}. Similarly, and although it
is an {\it excited} state of the system (bearing an
``anomalous'' or ``negative energy'' mode), the CV is generically
stable, independently of the chemical potential (or effectively
the number of atoms) of the system~\cite{middelpego}.


\begin{figure}
\begin{center}
    \includegraphics[width=7cm]{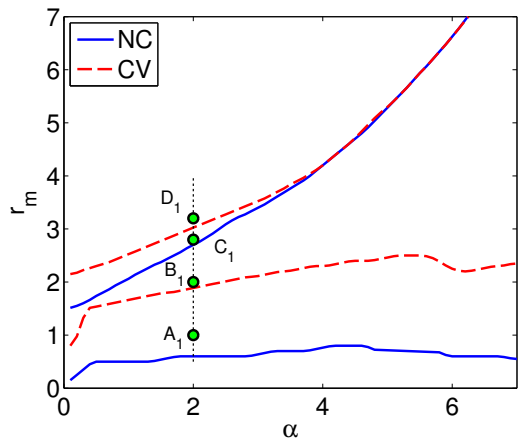}
\caption{(Color online)
Stability domains of nodeless cloud (NC) ,
and central vortex (CV) soliton-like solutions
for $\sigma=0.35$.
The stable domains correspond to the regions between
the curves.
Circles indicate points shown in following figures for
NC and CV solutions. 
 All quantities shown here and
in the figures that follow are dimensionless.
}
\label{fig:ranges1}
\end{center}
\end{figure}

\begin{figure}
\begin{center}
   \includegraphics[width=8.8cm]{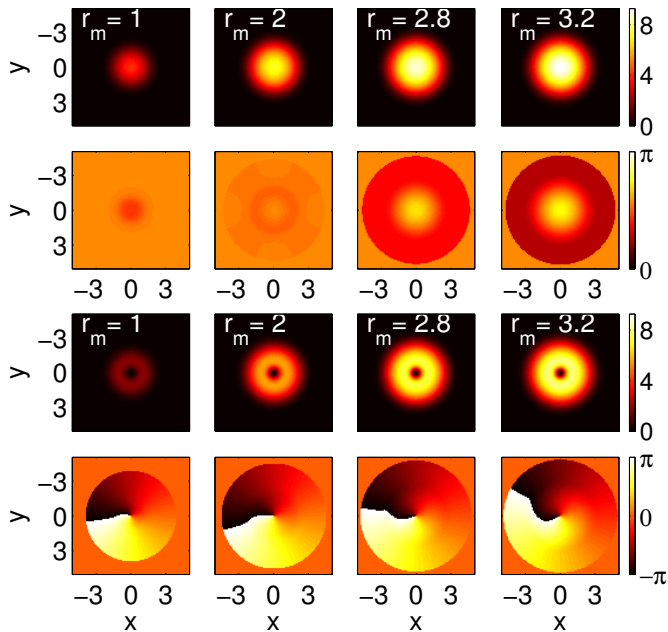}
\caption{(Color online) Density and phase profiles of
nodeless cloud configurations (NC, top two rows)
and central vortex configurations (CV, bottom two rows)
for $\alpha=2.0$ and $\sigma=0.35$.
The values of $\alpha$ and $r_m$ used correspond, left
to right, to points A$_1$-D$_1$ in Fig.~\ref{fig:ranges1}.
}
\label{fig:nc_prof}
\end{center}
\end{figure}

\begin{figure}
\begin{center}
    \includegraphics[width=8.8cm]{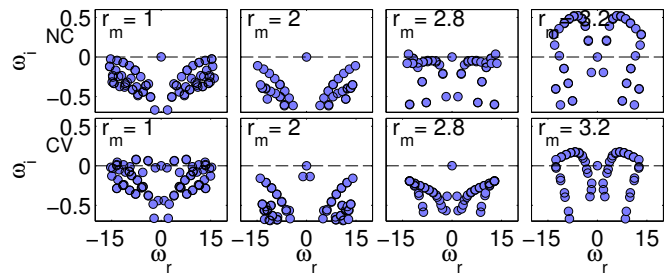}
\caption{(Color online) Eigenfrequencies ($\omega=\omega_r+i\,\omega_i$)
associated with the
spectral analysis of the NC (top) and CV (bottom) waveforms
for the same parameter values as
in Fig.~\ref{fig:nc_prof}.
}
\label{fig:spec_nccv} \end{center}
\end{figure}
The results of the scan of the parameter space are represented in
Fig.~\ref{fig:ranges1}, where we show the limits of stability of the NC and CV.
Interestingly, it can be observed that while there are
wide parametric regimes where the NC is stable, there are also
large intervals of parameters where this solution is, in fact,
unstable, contrary to what is known to be the
case in atomic BECs.
Furthermore, the stability region for the NC configuration is bounded
both from above and from below, unlike the one-dimensional
(1D) scenario of Ref.~\citenum{Cue11}
where the stability region is only bounded above.
Since there is loss everywhere, as the spotsize goes to zero there
is only enough gain to sustain an ever smaller condensate, until it
disappears in the limit of $r_m=0$.
On the other hand, also the CV features
wide intervals of stability, but also ones of instability.
It can, in fact, be seen that the feature identified as ``stability
inversion'' in Ref.~\citenum{Cue11} for the case of 1D polariton BECs is
still present here. Namely, there are regimes where the NC is
stable but the CV is not, but also ---in reverse--- there are regimes
where the CV is stable, but the NC is not.

In Fig.~\ref{fig:nc_prof} we
show the density and phase profiles of
the NC (top two rows) and CV (bottom two rows)
solutions for varying $r_m$. It is clear that as the radius of
the drive increases, so does the size of the condensate.
This is in contrast to what is the case with atomic BECs, where the size of the NC is controlled
solely by the (parabolic) trap: here,
the gain (and its interplay with the nonlinear loss/saturation) plays
a critical role in the size of the waveform.

For the CV solutions of the bottom two rows,
notice the characteristic $2\pi$ phase circulation.
Similar features to the NC case are observed
for the background on which the vortex ``lives''.

We now turn to an examination of the stability of the different
configurations. The spectral planes
$(\omega_r,\omega_i)$, where the subscripts $r$ and $i$ denote,
respectively the real and imaginary parts of the eigenfrequency,
for the NC and the CV configurations
are illustrated in Fig.~\ref{fig:spec_nccv}.
There, it is evident that except for a weak instability
arising through Hopf bifurcations for small values of $r_m$,
for most intermediate values
of $r_m$, both the NC and the CV configuration are stable.
The predominant instability that arises for both configurations
is the one for higher values of
the gain $r_m$ in this continuation. In that case, the instability arises in a
less customary (for such structures, at least in their Hamiltonian
form) way:
entire segments of the continuous spectrum cross
over the axis of $\omega_i=0$ (see right panels of Fig.~\ref{fig:spec_nccv})
and lead to bands of unstable
eigenfrequencies. It is, thus, in a sense, perhaps expected that the
entire ``background state'' of the system will be highly unstable
towards a fundamentally different pattern, an expectation that indeed
we will see to be confirmed by the direct numerical simulations
featuring the instability evolution of these states.

\subsubsection{Gray Rings and Triangular States}

\begin{figure}
\begin{center}
    \includegraphics[width=8.8cm]{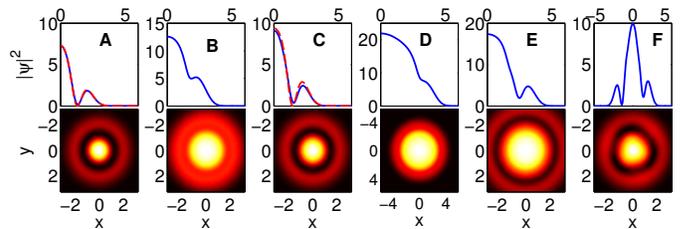}
\caption{(Color online) Profiles of gray rings (GR) ,
and triangular solutions (TS) for the system
parameters marked by (green) dots in the 2D plane of
Fig.~\ref{fig:dr_params}; $\sigma=0.35$. Dashed (red) lines represent
the Hamiltonian dark ring for the same chemical potential (panels A and C).
Panel F represents a TS solution.
}
\label{fig:dr_prof}
\end{center}
\end{figure}

\begin{figure}
\begin{center}
    \includegraphics[width=7.0cm]{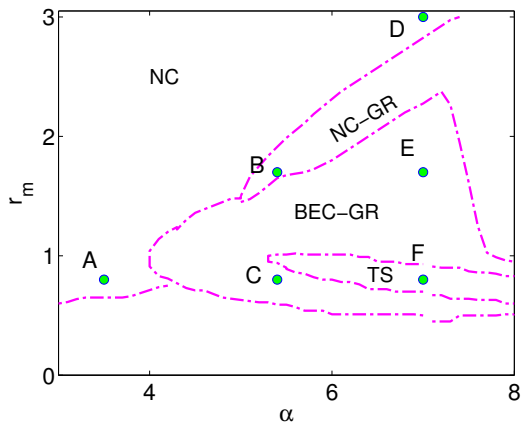}
\caption{(Color online) Existence domains of gray rings (GR),
and triangular solutions (TS). Solutions are stable except that
the GR is unstable where the TS exists.
for $\sigma=0.35$.
}
\label{fig:dr_params}
\end{center}
\end{figure}


In addition to the more fundamental solutions explored above, we have identified a host
of previously undisclosed, to the best of our knowledge, solutions of the 2D
polariton BEC system. Arguably, the most remarkable among them is the so-called gray ring (GR)
solitary wave, some typical density profiles of which are shown in
Fig.~\ref{fig:dr_prof}.
The depicted profiles correspond to the points shown as (green)
dots in the 2D existence/stability $(r_m,\alpha)$-plane (i.e., width and
amplitude of the parametric gain) shown in Fig.~\ref{fig:dr_params}.
The plane itself reveals some of the interesting
potential of such solutions, including the possibility of symmetry-breaking bifurcations
giving rise to a new triangular form of solutions (denoted as TS).
An additional remarkable feature is that while such ring dark solitons were proposed in
both atomic condensates~\cite{herring,todd,theoch_1} and in cGL gain/loss systems (as
radial Nozaki-Bekki holes) in both contexts they were found to be unstable and thus break up
into more prototypical coherent structures including vortices and spiral waves, respectively.
We now turn to a more detailed examination of their properties including the existence and
stability domains, also highlighting similarities and importantly differences from the above atomic
case, including the 
generically gray nature of such excitations in our polariton setting.

For relatively low values of the gain spot size ($r_m <1.3$) the GR
can be either stable ($\alpha> 4.0$) or unstable. Above that value, the
GR solution can
only be identified within the parametric range indicated by the curves in
Fig.~\ref{fig:dr_params}.
Interestingly, what is illustrated by the parametric plane is a progressive
convergence of the GR and NC solutions, as represented  by the label NC-GR
in Fig.~\ref{fig:dr_params}. This gradual ``merging'', at the solution
profile level, is signaled by the progressive
increase of the phase variation of the NC until it reaches a value around $\pi$.
Recall that from the results presented for the NC branch (Fig.~\ref{fig:nc_prof}), the phase
varies by less than $\pi/2$.
The admittedly somewhat arbitrary distinction between the NC branch and the GR one (and their hybrid
NC-GR form) herein was based on whether there
is a dip in the density profile (NC-GR) or not (NC).
In particular, we observe that as the value of the gain grows,
the NC becomes progressively more modulated.

On the other hand, there is an additional connection of such GR solutions with their
BEC analogs discussed earlier in Refs.~\citenum{herring,todd,theoch_1}. In particular, as
we approach the limit of weak and narrow drive ($\alpha \rightarrow 0$ and $r_m$ small),
the solution increasingly resembles the BEC  ring dark soliton (RDS) of the above works.
This is illustrated in panels A and C of Fig.~\ref{fig:dr_prof} where in addition to the polaritonic
GR profile, the
corresponding Hamiltonian case is also shown for the same chemical potential. As can be seen,
the dip in both cases occurs at the same position, that is distinct from the gain spot radius. This (and the overall quality of the density comparison)
is a strong indication of the origin of this solution from its corresponding Hamiltonian sibling.
Recalling that the RDS in the Hamiltonian case is always unstable, we identify herein
a critical role of  the gain along with the saturating loss terms in inducing a limited region
where this GR can be stabilized. As an addendum, it should be pointed out that the distinction
between accordingly termed BEC-GRs and the previously discussed NC-GRs is obtained through the
non-monotonic dependence of their dip versus the gain radius $r_m$.  More specifically, for BEC-GRs,
the depth of their dip (measured as the difference of the density at the center minus the density
at the dip) is found to increase with increasing $r_m$, while NC-GRs are instead characterized
by a decreasing dip (and NCs by a non-existent one). In reality, these solutions seem to seamlessly
merge as the critical points identified are traversed, however, the above distinctions were given
in order to better appreciate the ``origin'' of the different solutions.

An additional comment should be made here about the gray nature of these rings. Contrary to
what is expected from their BEC siblings featuring a phase shift of $\pi$ when stationary
(and associated with a finite velocity when they are ``gray''), in the polaritonic case, the
rings are generically found to be gray. This is, in fact, reminiscent of what was recently
found also in the context of complex $\mathcal{PT}$-symmetric potentials, e.g.,
in the work of Refs.~\citenum{vassos,konotopref}.  In both cases, the origin of the phenomenon is the same:
in particular, the complex nature (of the potential in the $\mathcal{PT}$-symmetric case
and of the gain/loss structure in the cGL setting herein) of the terms in the equation produce
a genuinely complex solution with a non-trivial phase structure and an associated ``particle flux''
along the stationary spatial profile. These features are absent in the BEC case, where the stationary RDS solution
is genuinely real.

In passing we note that another GR-like solution
we found, but that was never found to be stable, has its dip closer to the value
of the gain spot size. It is therefore related to this length scale, contrary to what is
the case for the GR
solutions focused upon here. Due to the generic instability of the waveform apparently
slaved to the gain, we do not explore it further here.

\begin{figure}
\begin{center}
    \includegraphics[width=8.8cm]{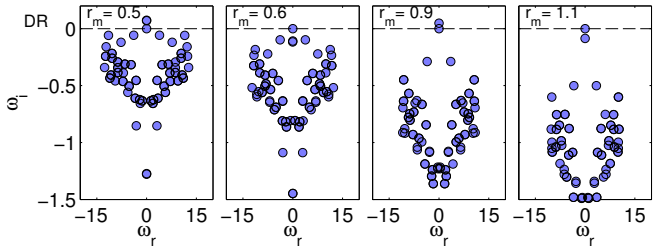}
\caption{(Color online) Spectra pertaining to the GR
solution for  $\alpha=6.0$. Notice that for $r_m=0.9$ the unstable GR solution
cohexists with a stable TS.
}.
\label{fig:spec_dr}
\end{center}
\end{figure}

The spectral properties of the GR state have been found to be significantly different than those of
the previous two fundamental states (NC and CV). In particular, as can be seen in
Fig.~\ref{fig:spec_dr}, the instability of this state as $r_m$
is increased stems from the fact that
eigenfrequencies cross
the origin of the spectral plane.
This predisposes us towards a fundamentally different instability, possibly arising through
a symmetry breaking pitchfork bifurcation. We will see also how this
expectation is manifested in the direct numerical simulations of
the following section.
This symmetry breaking bifurcation
is substantiated in the parametric plane of Fig.~\ref{fig:dr_params} where a region
has been denoted under TS (triangular solutions). As the curve outlining this region is
crossed, the BEC-GR solutions undergo the above mentioned pitchfork bifurcation and
spontaneously give rise to such TS. The TS states are generically found to be stable,
a feature that will render them natural attractors for the BEC-GR unstable dynamics in the parametric
range of TS existence, as will be corroborated in the dynamical evolution section below.

\subsubsection{Quadrupoles}

In the same spirit as the triangular solutions identified above, we have also been able
to find solutions with quadrupolar symmetry.
These solutions are especially relevant below the region of stability of the
NC, as well as that of the (BEC-)GR.
Such solutions are characterized by four dips of the density, at which location the phase
portrait shows a winding of $2\pi$ (Fig.~\ref{fig:qs_prof}).
This type of solution is also reminiscent of a corresponding quadrupole (vortex) solution
in the realm of atomic BECs~\cite{middelpego}. In the latter setting, the solutions are
critically induced by the parabolic trap, created from the linear limit thereof (as a complex
combination of two of the second excited states of the 2D quantum-harmonic oscillator).

\begin{figure}
\begin{center}
    \includegraphics[width=7.0cm]{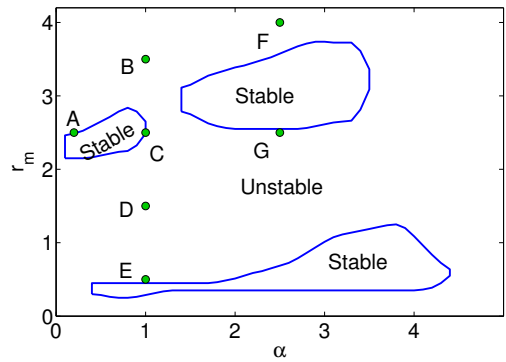}
\caption{(Color online) Stability domains of quadrupole solutions (QS),
for $\sigma=0.35$. The different letters denote case example
profiles illustrated below in Fig.~\ref{fig:qs_prof}.
}
\label{fig:qs_params}
\end{center}
\end{figure}

\begin{figure}
\begin{center}
    \includegraphics[width=8.6cm]{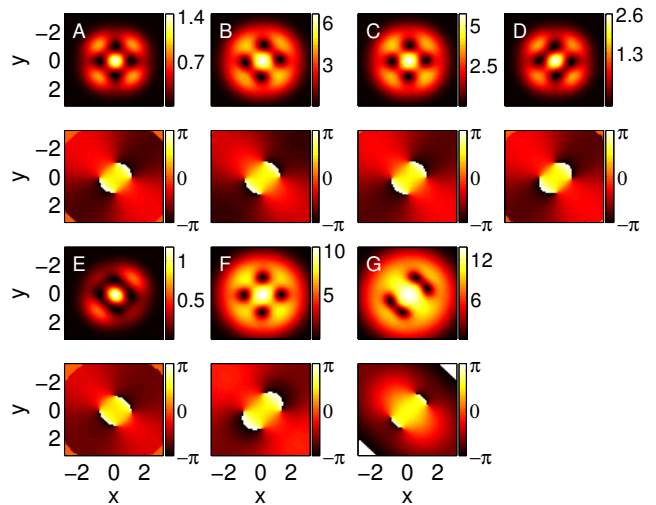}
\caption{(Color online) Density and phase profiles of quadrupole solutions (QS) for the values of
$\alpha$ and $r_m$ corresponding  to
points A-G in Fig.~\ref{fig:qs_params}.
}
\label{fig:qs_prof}
\end{center}
\end{figure}

The regions of
stability of the quadrupoles in the two parameter plane of $(\alpha,r_m)$
are shown in Fig.~\ref{fig:qs_params}. Different classes of instability
can be found, leading to exponential, oscillatory or combined
decay. Nevertheless, islands of stability are also identified within which
as we will see below the quadrupolar state can offer a dynamical attractor
starting, e.g., from GR initial data, but also for both the NC and the CV.
%

As the gain strength grows, we find that the quadrupole exact solution profiles  tend
 to a more elongated profile along a symmetry axis,
reaching a form where the four dips nearly coalesce in two.
Nevertheless, the phase (and the vorticity, not shown) reveal the four vortices
of alternating positive and negative charge are still separate.

\subsection{Dynamical evolution}

As a way of confirming the stability results found above,
and also of exploring the pattern forming outcomes of the dynamical
evolution of the identified instabilities,
we numerically integrated the full
equation of motion, namely Eq.~(\ref{eq:dyn}). Our initial conditions
consisted of profiles in the form
of the above obtained exact (up to a prescribed numerical tolerance)
solutions, suitably perturbed to accelerate the decay, if unstable, or
confirm that it returns to the attracting solution, if stable. The perturbation
added for unstable waveforms
was in the form of the profile of the eigenvector with the most
unstable eigenfrequency. For the solutions expected to be stable,
random noise was used. The latter results are not shown (they were
only used to confirm the spectral stability results), but
it was
found that the solutions
remained unaltered after propagation for time up to $t=1000$.

\begin{figure}
\begin{center}
    \includegraphics[width=8.8cm]{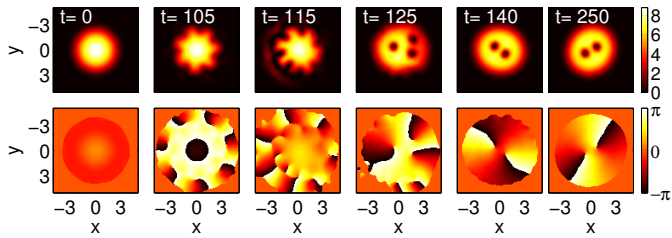}
\caption{(Color online) Dynamical evolution of a NC state for parameter values $\alpha=2.0$ and $r_m=2.8 $,
which is just outside of the stability region.
}
\label{fig:dynNC_instC}
\end{center}
\end{figure}

\begin{figure}
\begin{center}
    \includegraphics[width=8.8cm]{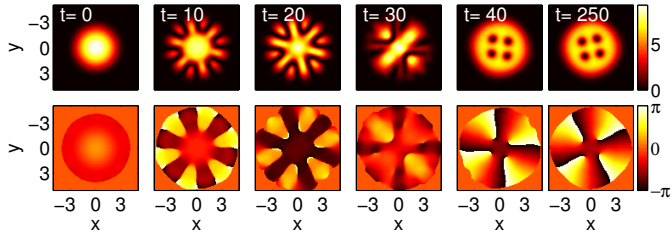}
\caption{(Color online) Dynamical evolution of a NC state for parameter values $\alpha=2.0$ and $r_m=3.2 $. The resulting configuration of 4 vortices is rotating.
}
\label{fig:dynNC_instD}
\end{center}
\end{figure}

Examples of the unstable scenarios are presented below. The results presented
correspond again to points of the set A$_1$-D$_1$ that are unstable (for each
of NC and/or CV);
other values lead to a qualitatively similar behavior.
Figure~\ref{fig:dynNC_instC}
illustrates the case where the NC has just become unstable (for $r_m=2.8$, while the
stability boundary for this value of $\alpha=2$ is $r_m=2.7$). In this
case, we can observe that the NC becomes unstable to an azimuthal
modulation with eight-fold symmetry

and eventually, upon the nonlinear
evolution of the instability, decays to a pair of rotating vortices. A further
increase of $r_m=3.2$ results in a similar evolution
(cf.~Fig.~\ref{fig:dynNC_instD}), but
the end result is a rotating lattice of four vortices.
Here we see a different initial symmetry in the dominant unstable
mode (which appears to create a hexagonal
modulation; see, e.g., the snapshots at $t=10$
and $t=20$), but also a nonlinear intermediate step in the evolution (see, e.g., snapshot at $t=30$).
It is naturally expected that, as $r_m$ increases and the polariton condensate
accordingly grows, more vortices can be ``accommodated'' therein, i.e.,
brought in the domain from the outside, and hence larger lattices created; this is
in line also with the observations of Ref.~\citenum{berloff1}.
The evolution typically starts with a modulation around the edge of the cloud
that starts rotating. Comparison with the profile of the most unstable
eigenvector suggests that the latter is indeed
responsible for this (increasing) modulation.
It eventually leads to one or more vortices spiralling (from the periphery)
to the central region of the cloud until a stable and symmetric arrangement
of vortices is achieved. The number of vortices (in the parameter range
studied) can vary from one up to 21, either in a ring shape, with one
more at the center, or as a lattice when their number grows.
%
%

\begin{figure}
\begin{center}
    \includegraphics[width=8.8cm]{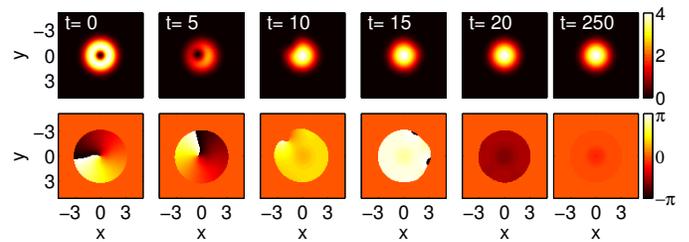}
\caption{(Color online) Dynamical evolution of a CV state for parameter
values $\alpha=2.0$ and $r_m=1.0 $, which is below its stability region.
}
\label{fig:dynCV_instA}
\end{center}
\end{figure}

\begin{figure}
\begin{center}
    \includegraphics[width=8.8cm]{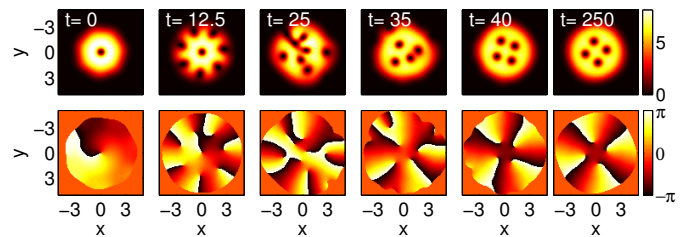}
\caption{(Color online) Dynamical evolution of a CV state for parameter
values $\alpha=2.0$ and  $r_m=3.2 $, which is just outside (above) its stability region.
The resulting configuration appears to 
ultimately lead to a rotating square (cf. Fig.~\ref{fig:dynNC_instD}
for the same parameter values).
}
\label{fig:dynCV_instD}
\end{center}
\end{figure}

The evolution of the CV solution
features a similar behavior to the NC for parametric
values beyond the upper stability boundary of this solution.
In the region below the lower stability boundary, where the NC is generically stable, the CV
 typically decays to the NC (see, e.g., Fig.~\ref{fig:dynCV_instA}).
Above the higher stability boundary of the CV solution, the vortex
decays to a lattice of vortices as shown in Fig.~\ref{fig:dynCV_instD}.
It is interesting to note that
the original evolution of the instability results in
more vortices within the cloud than the resulting asymptotic state,
so there is a ``distilling'' process taking place, which finally results
in the rotating square configuration observed at longer times
(cf.~for the same parameters, the asymptotically
favored configuration of Fig~\ref{fig:dynNC_instD}).
%

\begin{figure}
\begin{center}
    \includegraphics[width=8.8cm]{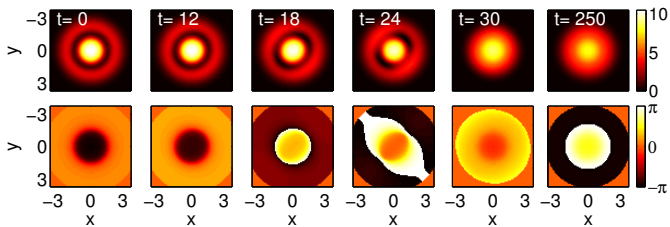}
\caption{(Color online) Dynamical evolution of a GR state for parameter values $\alpha=4.2$ and $r_m=1.3 $, which is above its stability region.
The final excitation is a NC state.
}
\label{fig:dynDR_instE1}
\end{center}
\end{figure}

\begin{figure}
\begin{center}
    \includegraphics[width=8.6cm]{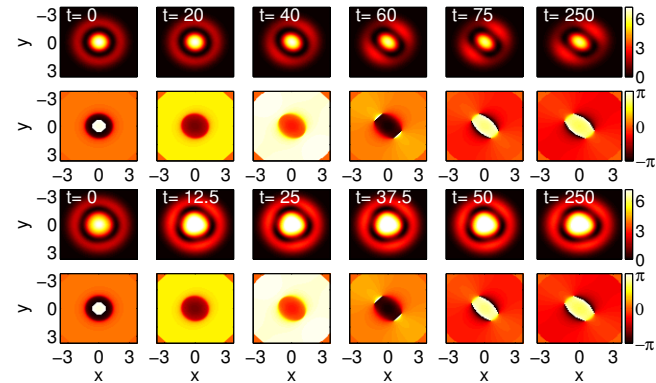}
\caption{(Color online) The top two rows show the
dynamical evolution of a GR state for parameter values
$\alpha=4.2$ and $r_m=0.7$, which is below its stability region.
The final excitation features axial symmetry and is a QS.
The bottom two rows show the dynamical evolution of a GR state for
parameter values $\alpha=6.9$ and  $r_m=0.8 $, which
is also outside its stability region.
Here, however, the final excitation features triangular symmetry (TS).
}
\label{fig:dynDR_instE2}
\end{center}
\end{figure}


We now turn to the dynamical results for the evolution of the GR
solutions.
Above its stability region, the GR typically decays towards the NC, which
for this parametric region is well within its own range of stability in the
$(\alpha,r_m)$ plane; this is shown, e.g., in Fig.~\ref{fig:dynDR_instE1}, for $\alpha=4.2$ and $r_m=1.3$. In contrast,
below the lower stability threshold of the GR, the latter
may decay to different solutions depending on the exact parameters
(and perturbations) used.  An example of a relevant
possibility is shown in the top two rows of Fig.~\ref{fig:dynDR_instE2}
for $\alpha=4.2$ and $r_m=0.7$. This case reveals the possibility
that the symmetry breaking instability of the GR (cf. the discussion
of the previous section) may result into a quadrupolar
configuration of the type explored in the previous section.
On the other hand, the bottom two rows
of Fig.~\ref{fig:dynDR_instE2} illustrate a different
scenario for the parameter
set $\alpha=6.9$ and $r_m=0.8$, which can be identified as being
within the region of the symmetry breaking instability towards triangular solutions.
In particular, the symmetry breaking spontaneously manifests itself dynamically
resulting towards a configuration with triangular symmetry, as may be expected based
on our existence/stability earlier findings.
Solutions with this symmetry were previously identified in other
contexts, e.g., Ref.~\citenum{BorLob12}.
Nevertheless, these results, and the absence of vortex lattice formation in this case,
are in accordance
with the results of the stability analysis of the
GR presented
above and its fundamentally different instability
mechanism in comparison to the NC or CV configurations.

\begin{figure}
\begin{center}
    \includegraphics[width=8.8cm]{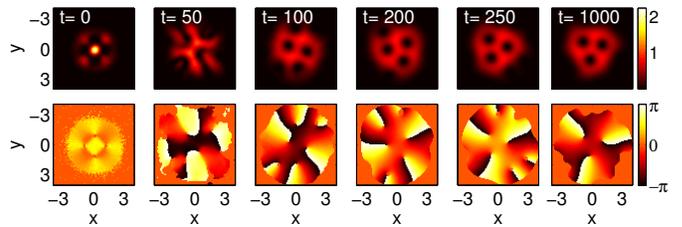}
\caption{(Color online)
Dynamical evolution of a QS for parameter values $\alpha=0.2$ and $r_m=2.5$
(see point labeled A in Fig.~\ref{fig:dynNC_instC}).
The final state is a rotating lattice of 3 vortices.
}
\label{fig:dynQS_A}
\end{center}
\end{figure}

\begin{figure}
\begin{center}
    \includegraphics[width=8.8cm]{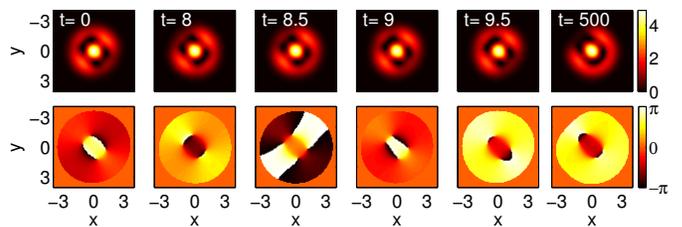}
\caption{(Color online)
Dynamical evolution of a QS for parameter values $\alpha=2.5$ and  $r_m=0.9$ 
(not in  Fig.~\ref{fig:dynNC_instC}). It corresponds to an oscillatory instability.
}
\label{fig:dynQS_osc}
\end{center}
\end{figure}

\begin{figure}
\begin{center}
    \includegraphics[width=8.8cm]{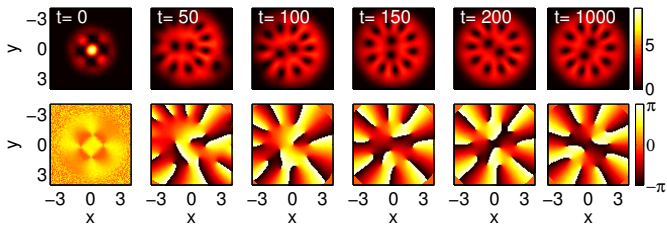}
\caption{(Color online)
Dynamical evolution of a QS for parameter values $\alpha=1.0$ and  $r_m=3.5$
(see point labeled B in Fig.~\ref{fig:qs_params}).
The final state is a rotating lattice of
10 vortices in a necklace plus 2 at the center.
The two sets rotate at different rates. (Initial frame has its colormap scaled
by a factor of 1/5).
}
\label{fig:dynQS_B}
\end{center}
\end{figure}

\begin{figure}
\begin{center}
    \includegraphics[width=8.8cm]{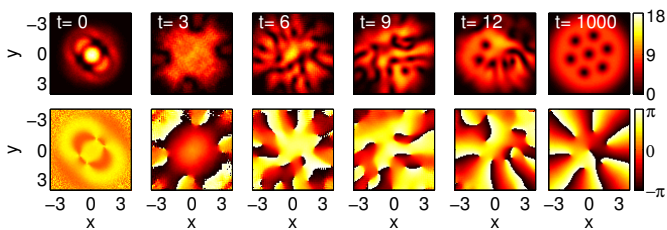}
\caption{(Color online)
Dynamical evolution of a QS for parameter values $\alpha=2.5$ and $r_m=4.0$
(see point labeled F in Fig.~\ref{fig:qs_params}).
The final state is a (slowly) rotating lattice of 7 vortices.
(Initial frame has its colormap scaled by a factor of 1/10).
}
\label{fig:dynQS_F}
\end{center}
\end{figure}

The evolution of the QS solutions shows a range of behaviors, from a decay towards a NC
(for points D and G in Fig~\ref{fig:qs_params}) to rotating lattices of vortices
(A, B, C, and F). Among these, we highlight, in particular, the scenario A ($\alpha=0.2$, $r_m=2.5$)
, shown in Fig~\ref{fig:dynQS_A}, where the final rotating cloud is highly distorted,
as if more vortices were trying to join the 3 already in the central region of the cloud.
It can also display an oscillatory instability, as in Fig~\ref{fig:dynQS_osc}, where the two
dips at the extrema of the axial central lobe perform an oscillation, each pair with
opposite phase than the other.
Yet another  situation is exemplified in Fig~\ref{fig:dynQS_B},
where a lattice of vortices results, but unlike all other so far seen
the inner vortices rotate at a different rate than the outer ones.
Finally, the case F (see Fig.~\ref{fig:dynQS_F})
results in a final excitation where a lattice
with 7 vortices (one at the centre and an hexagon of vortices around
it) rotates very slowly.


\section{Conclusions}

In the present work, motivated by the intensely studied theme of
polariton condensates, we offered a detailed view of the existence
and stability, as well as the nonlinear dynamical
properties of some prototypical states appearing in such systems.
These states included fundamental earlier revealed waveforms such as
the nodeless cloud (NC) and the cloud with a central vortex (CV).
For these, we presented a systematic two-parametric analysis of their stability
properties and how these are reflected in the corresponding nonlinear dynamics.

The fundamental (especially in the atomic BEC case)
nodeless state was found to be stable only in a limited range of parameters.
The most simple excited state, the central vortex state (which is again generically
robust in atomic BECs), was found to, in fact, potentially
exist as a stable object for parameters where the NC is no longer stable. Vice
versa, the nodeless cloud was also stable in regimes where the vortex
was not, presenting a 2D generalization of the stability
inversions reported earlier in 1D counterparts of the model \cite{Cue11}.
Outside their stability ranges, they were both found to decay
towards a series of rotating vortex lattices, in line with earlier
numerical observations
\cite{berloff1}. However, here the precise (Hopf) nature of the instability
was elucidated and the unusual associated
morphology featuring the destabilization of an entire band of continuous
spectrum eigenfrequencies was revealed.

As regards vortices, it should also be noted that we attempted
to identify doubly charged vortex solutions, however, we were unable
to obtain spectrally stable such structures in the realm of the
present model, i.e., such states were identified yet were always
found to be dynamically unstable.

In addition to these simpler structures, we also explored more elaborate
ones, especially in the form of
a gray ring (GR) soliton-like structure, which was connected both to the NC
but also to the ring dark soliton (RDS) state of atomic BECs. The inclusion of
gain and loss in our complex Ginzburg-Landau-like  equation was found to play
a critical role in the potential stabilization of such ring states. Their generic
gray structure was also justified by the flux induced by the gain/loss.
Aside from identifying the stability islands of such GR states, their potential
dynamical instabilities and associated bifurcations were also revealed.
These were shown to lead to symmetry-breaking events generating (and asymptoting to) solutions
of potentially triangular or quadrupolar structure. These states, in turn,
were also identified as exact stationary solutions and their own two-parametric
stability properties were explored.

A different set of gray ring structures was found but never stable, from what we
could determine. This solution is mainly characterized by a central peak lower than
the outer ring. It exists in regions in parameter space where other solutions exist stably.
Thus for certain regions of parameters it is possible to find a NC, a GR and 
this other, always unstable
gray ring.

All the results reported were for the parameter $\sigma=0.35$, as stated before.
Both from our previous results in the 1D setting, and from intuition
and case examples considered, we expect that
extending the continuation to other (nearby) values of  $\sigma$ should yield
qualitatively similar results.

There are
many directions towards which this exploration
could further proceed. On the one hand, in the setting with the
parabolic trap, it is interesting to explore the detailed stability of vortex clusters and progressively
growing configurations towards vortex lattices. Understanding these
clusters is still a very active area of research in atomic BECs~\cite{ourprl};
extending relevant (vortex) particle approaches (or distributional
ones~\cite{anglin}) in order to understand the properties
and internal modes (in analogy to the Tkachenko modes
of atomic BECs~\cite{engels}) of the
system of a few or of many vortices, would be of particular interest
in its own right. Yet another interesting direction, given the significant
progress in imposing potentials of different kinds including periodic
ones and identifying states critically supported by them (such as
the gap solitons of Ref.~\citenum{amogap}), would be to explore the interplay
of such clusters and lattices with external potentials and their
structural phase transitions between different energetically preferred
states as, e.g., the lattice parameters of an external
periodic potential are varied~\cite{pu}. These and other
related topics are presently under consideration and will be reported
in future publications.

\section*{Acknowledgments}
P.G.K. gratefully acknowledges support from the National Science Foundation,
under grant CMMI-1000337, from the AFOSR under grant FA9550-12-1-0332,
from the Binational Science Foundation under grant 2010239 and finally
the warm hospitality of the IMA of the University of Minnesota 
and of KIP and the University of Heidelberg during the
final stages of this work. A.S.R. also acknowledges the hospitality of
KIP and of the University of Heidelberg.
The work of D.J.F. was partially supported by
the Special Account for Research Grants of the University of Athens. J.C. and F.P. acknowledge financial support from the
MICINN project FIS2008-04848.


\begin{thebibliography}{99}

\bibitem{kasp1_and_more}
J. Kasprzak, M. Richard, S. Kundermann, A. Baas, P. Jeambrun,
J. Keeling, F.M. Marchetti, M.H. Szyma\'nska, R. Andr\'e,
J.L. Staehli, V. Savona, P.B. Littlewood, B. Deveaud, and L.S. Dang,
Nature {\bf 443}, 409 (2006).
%

\bibitem{balili} R. Balili, V. Hartwell, D. Snoke, L. Pfeiffer, and K. West,
Science {\bf 316}, 1007 (2007).
%

\bibitem{lai} W. Lai, N.Y. Kim, S. Utsunomiya, G. Roumpos, H. Deng, M.D. Fraser,
T. Byrnes, P. Recher, N. Kumada, T. Fujisawa, and Y. Yamamoto,
Nature {\bf 450}, 529 (2007).
%

\bibitem{deng} H. Deng, G.S. Solomon, R. Hey, K.H. Ploog, and Y. Yamamoto,
Phys.\ Rev.\ Lett.\ {\bf 99}, 126403 (2007).


\bibitem{benoit} B. Deveaud (Ed.), {\it The Physics of Semiconductor
Microcavities} (Wiley-VCH, Weinheim, 2007).

\bibitem{sanvitto} {\it Exciton Polaritons in Microcavities, New Frontiers},
Springer Series in Solid-State Sciences Vol. 172, edited
by D. Sanvitto and V. Timofeev (Springer, New York,
2012).

\bibitem{rmp1} H. Deng, H. Haug, and Y. Yamamoto, Rev. Mod. Phys. {\bf 82}, 1489 (2010).

\bibitem{berloff_review} J. Keeling and N.G. Berloff,
Contemporary Phys. {\bf 52}, 131 (2011).

\bibitem{rmp2} I. Carusotto and C. Ciuti, Rev. Mod. Phys. {\bf 85}, 299 (2013).

\bibitem{amo1} A. Amo,
J. Lefr{\`e}re, S. Pigeon, C. Abrados, C. Ciuti,
I. Carusotto, R. Houdr{\'e}, E. Giacobino, and A. Bramati,
Nature Phys. {\bf 5}, 805 (2009).

\bibitem{lagou1}
K.G. Lagoudakis,
M. Wouters, M. Richard, A. Baas,
I. Carusotto, R. Andr{\'e}, L.S. Dang, and B. Deveaud-Pl{\'e}dran,
Nature Phys. {\bf 4}, 706 (2008); see also for
half-quantum vortices the work of
K.G. Lagoudakis,
    T. Ostatnick{\'y},
    A.V. Kavokin,
    Y.G. Rubo,
    R. Andr{\'e},
    B. Deveaud-Pl{\'e}dran, Science {\bf 326}, 974 (2009).


\bibitem{roumpos} M.D. Fraser, G. Roumpos, and Y. Yamamoto,
New J.\ Phys.\ {\bf 11}, 113048 (2009).

\bibitem{roumpos1} G. Roumpos, M.D. Fraser, A. Loffler, S. H{\"o}ffling,
A. Forchel, Y. Yamamoto,
Nature Phys. {\bf 7}, 129 (2011);
G. Nardin, G. Grosso, Y. L{\'e}ger, B. Pietka, F. Morier-Genoud
and B. Deveaud-Pl{\'e}dran {\bf 7}, 635 (2011);
G. Tosi, F.M. Marchetti, D. Sanvitto, C. Ant{\'o}n, M.H. Szyma{\'n}ska,
A. Berceanu, C. Tejedor, L. Marrucci, A. Lema\^{\i}tre, J. Bloch,
and L. Vi\~na, Phys. Rev. Lett. {\bf 107}, 036401 (2011).

\bibitem{sanvitto_nat}
D. Sanvitto, F.M. Marchetti, M.H. Szyma{\'n}ska, G. Tosi, M. Baudisch,
F.P. Laussy, D.N. Krizhanovskii, M.S. Skolnick, L. Marrucci, A. Lema\^{i}tre,
J. Bloch, C. Tejedor, and L. Vi\~na,
Nature Phys. {\bf 6}, 527 (2010).

\bibitem{amo2}
A. Amo, D. Sanvitto, F.P. Laussy, D. Ballarini,
E. del Valle, M.D. Martin, A. Lema\^{\i}tre, J. Bloch,
D.N. Krizhanovskii, M.S. Skolnick, C. Tejedor, and L Vi\~na,
Nature {\bf 457}, 291 (2009).


\bibitem{skryabin} M. Sich, D.N. Krizhanovskii, M.S. Skolnick, A.V. Gorbach,
R. Hartley, D.V. Skryabin, E.A. Cerda-M{\'e}ndez, K. Biermann,
R. Hey, and P.V. Santos, Nature Photonics {\bf 6}, 50 (2012).

\bibitem{dev} G. Grosso,G. Nardin, F. Morier-Genoud,Y. L{\'e}ger, and B.
Deveaud-Pl{\'e}dran, Phys. Rev. Lett. {\bf 107}, 245301 (2011).

\bibitem{amogap} D. Tanese, H. Flayac, D. Solnyshkov, A. Amo,
A. Lema\^{i}tre, E. Galopin, R. Braive, P. Senellart, I.
Sagnes, G. Malpuech, and J. Bloch, Nat. Commun. {\bf 4}, 1749 (2013);
see also for a theoretical
analysis of gap solitons in polariton condensates the recent
work of E.A. Ostrovskaya, J. Abdullaev, M.D. Fraser,
A.S. Desyatnikov, and Yu.S. Kivshar
Phys. Rev. Lett. {\bf 110}, 170407 (2013).

\bibitem{amo3}
A. Amo, T.C.H. Liew, C. Adrados, R. Houdr{\'e},
E. Giacobino, A.V. Kavokin, and A. Bramati,
Nature Photonics {\bf 4}, 361 (2010).

\bibitem{amo4}
S.I. Tsintzos,
N.T. Pelekanos, G. Konstantinidis,
Z. Hatzopoulos, and P.G. Savvidis,
Nature {\bf 453}, 372 (2008).

\bibitem{polar1}
M. Wouters and I. Carusotto,
Phys.\ Rev.\ Lett.\ {\bf 99}, 140402 (2007).

\bibitem{polar2}
M. Wouters, I. Carusotto, and C. Ciuti,
Phys.\ Rev.\ B {\bf 77}, 115340 (2008).

\bibitem{cc05} C. Ciuti and I. Carusotto,
Phys.\ Stat.\ Sol.\ (b) {\bf 242}, 2224 (2005).

\bibitem{berloff1}
J. Keeling and N.G. Berloff,
Phys.\ Rev.\ Lett.\ {\bf 100}, 250401 (2008).

\bibitem{kbb}
M.O. Borgh, J. Keeling, and N.G. Berloff,
Phys.\ Rev.\ B {\bf 81}, 235302 (2010).

\bibitem{kbb2} M.O. Borgh, G. Franchetti, J. Keeling, and
N.G. Berloff, Phys. Rev. B {\bf 86}, 035307 (2012).

\bibitem{nbb_nature} G. Tosi G. Christmann, N.G. Berloff, P. Tsotsis,
T. Gao, Z. Hatzopoulos, P.G. Savvidis, J.J. Baumberg,
Nature Phys. {\bf 8}, 190 (2012).

\bibitem{dark_djf} D.J. Frantzeskakis, J. Phys. A {\bf 43}, 213001 (2010).

\bibitem{herring} G. Herring,
L.D. Carr, R. Carretero-Gonz\'alez, P.G. Kevrekidis, and D.J. Frantzeskakis,
Phys. Rev. A {\bf 77}, 023625 (2008).

\bibitem{todd} T. Kapitula, P.G. Kevrekidis, and R. Carretero-Gonz{\'a}lez,
Physica D {\bf 233}, 112 (2007).

\bibitem{theoch_1} G. Theocharis, D.J. Frantzeskakis, P.G. Kevrekidis,
B.A. Malomed, and Yu.S. Kivshar, Phys. Rev. Lett. {\bf 90}, 120403 (2003);
G. Theocharis, P. Schmelcher, M.K. Oberthaler, P.G. Kevrekidis, and D.J. Frantzeskakis,
Phys. Rev. A 72, 023609 (2005).

\bibitem{nozakibekki} K. Nozaki and N. Bekki, J. Phys. Soc. Jpn. {\bf 53}, 1581 (1984);
Phys. Lett. A {\bf 110}, 133 (1985).

\bibitem{rabin1} M. Bazhenov and M. Rabinovich, Phys. Lett. A {\bf 179}, 191 (1993).
M. Bazhenov, M. Rabinovich, Physica D {\bf 73}, 318 (1994).


\bibitem{OstAbd12}
E.A. Ostrovskaya, J. Abdullaev, A.S. Desyatnikov, M.D. Fraser, and Yu.S. Kivshar,
Phys. Rev. A {\bf 86}, 013636 (2012).


\bibitem{stringari1}
C.J. Pethick and H. Smith,
{\it Bose-Einstein condensation in dilute gases} (Cambridge University
Press, Cambridge, 2002).

\bibitem{stringari2}  L.P. Pitaevskii and S. Stringari,
{\it Bose-Einstein Condensation} (Oxford University Press, Oxford, 2003).

\bibitem{boris1} C.-K. Lam, B.A. Malomed, K.W. Chow, and P.K.A. Wai,
Eur. Phys. J. Special Topics {\bf 173}, 233
(2009);
%
C.H. Tsang, B.A. Malomed, C.-K. Lam, and K.W. Chow,
Eur. Phys. J. D {\bf 59}, 81
(2010);
%
F. Kh. Abdullaev, V.V. Konotop, M. Salerno, and A.V. Yulin,
Phys. Rev. E {\bf 82}, 056606 (2010);
%
V. Skarka, N.B. Aleksi\'c, H. Leblond, B.A. Malomed, and D. Mihalache,
Phys. Rev. Lett. {\bf 105}, 213901 (2010);
%
Y.V. Kartashov, V.V. Konotop, and V.A. Vysloukh, Europhys. Lett. {\bf 91}, 34003 (2010);
%
M.J. Ablowitz, T.P. Horikis, S.D. Nixon, and D.J. Frantzeskakis, Opt. Lett. {\bf 36}, 793 (2011);
%
Y.V. Kartashov, V.V. Konotop, and V.A. Vysloukh,
Opt. Lett. {\bf 36}, 82 (2011);
%
D.A. Zezyulin, Y.V. Kartashov, and V.V. Konotop,
Opt. Lett. {\bf 36}, 1200 (2011);
%
M.J. Ablowitz, S.D. Nixon, T.P. Horikis, and D.J. Frantzeskakis,
Proc. Roy. Soc. London A {\bf 467}, 2597 (2011);
%
Y.V. Kartashov, V.V. Konotop, and V.A. Vysloukh,
Phys. Rev. A {\bf 83}, 041806(R) (2011);
%
M.J. Ablowitz, S.D. Nixon, T.P. Horikis, and D.J. Frantzeskakis,
J. Phys. A {\bf 46}, 095201 (2013).


\bibitem{boris2}
B.A. Malomed, O. Dzyapko, V.E. Demidov, and S.O. Demokritov,
Phys. Rev. B {\bf 81}, 024418 (2010).
%




\bibitem{Cue11} J.\ Cuevas, A.S. Rodrigues, R. Carretero-Gonz{\'a}lez, P.G. Kevrekidis, and D.J. Frantzeskakis,
Phys. Rev. B \textbf{83}, 245140 (2011).

\bibitem{middelpego} S. Middelkamp, P.G. Kevrekidis, D.J. Frantzeskakis, R. Carretero-Gonz{\'a}lez, and P. Schmelcher
Phys. Rev. A {\bf 82}, 013646 (2010); see also
R. Koll{\'a}r and R.L. Pego, Appl. Math. Res. Express {\bf 1}, 1
(2012).

\bibitem{vassos} V. Achilleos, P. G. Kevrekidis, D. J. Frantzeskakis, and R. Carretero-González
Phys. Rev. A {\bf 86}, 013808 (2012).

\bibitem{konotopref} D.A. Zezyulin and V.V. Konotop
Phys. Rev. A 85, 043840 (2012).


\bibitem{BorLob12}
O.V. Borovkova, V.E. Lobanov, Y.V. Kartashov, and L. Torner,
Opt. Lett. {\bf 36}, 1936 (2011).


\bibitem{ourprl}
R. Navarro, R. Carretero-Gonz{\'a}lez, P.J. Torres, P.G. Kevrekidis,
D.J. Frantzeskakis, M.W. Ray, E. Altunta\c{s}, and D.S. Hall
Phys. Rev. Lett. {\bf 110}, 225301 (2013).

\bibitem{anglin} J.R. Anglin and M. Crescimanno,
arXiv:cond-mat/0210063.

\bibitem{engels}  I. Coddington, P. Engels, V. Schweikhard, and E.A. Cornell,
Phys. Rev. Lett. {\bf 91}, 100402 (2003).

\bibitem{pu} H. Pu, L.O. Baksmaty, S. Yi, and N.P. Bigelow,
Phys. Rev. Lett. {\bf 94}, 190401 (2005).


\end{thebibliography}
\end{document}